\documentclass[aps,pra,reprint,showkeys,longbibliography]{revtex4-2} 
\usepackage{amsmath}
\usepackage{amssymb}
\usepackage{graphicx} 
\usepackage{bm} 
\usepackage{xcolor}
\usepackage[toc,page]{appendix}
\usepackage{braket}
\usepackage{mathrsfs}

\makeatletter

\@ifclasswith{revtex4-2}{draft}{%
    \usepackage[notref,notcite]{showkeys}
    
}{}

\makeatother

\newcommand{\nn}{\nonumber}

\begin{document}

\title{Quantum mechanics for classical transport equations}

\author{Christof Wetterich}
\affiliation{Institut f\"ur Theoretische Physik\\
    Universit\"at Heidelberg\\
    Philosophenweg 16, D-69120 Heidelberg}

\begin{abstract}
    Classical transport equations with probabilistic initial conditions can be viewed as quantum systems.
    In a discrete version they are probabilistic automata.
    The time-local probabilistic information is encoded in a classical wave function.
    Its unitary evolution obeys a Schr\"odinger equation.
    Statistical observables measure properties of the classical probability distribution.
    In the quantum formalism they are represented by operators which do not commute with the ones associated to classical observables.
    Momentum, quantum energy and angular momentum or charges yield conserved quantities which constrain the evolution of the classical probability distribution.
    The characteristic features of quantum mechanics, as the superposition of wave functions, interference, the importance of phases, non-commuting operators or a unitary time evolution, are realized by probabilistic classical transport equations. 
    Stochastic transport equations are described by a quantum system with a stochastic Hamiltonian.
\end{abstract}

\maketitle

\section{Introduction}

\indent We consider probabilistic transport equations.
They involve a deterministic transport of some variables or fields $\sigma$ with time.
The initial conditions are given by a probability distribution for $\sigma$.
The central quantity is the time-dependent probability distribution $w (t,\sigma)$.
By use of a real classical wave functional $q(t,\sigma)$ as the root of the probability distribution, $w(t,\sigma)=q^2(t,\sigma)$, we show that this classical probabilistic system can be viewed as a quantum system.
Deterministic transport equations with probabilistic initial conditions are used in a wide area of physics and more general sciences. 
Examples range from fluid dynamics in cosmology, turbulence or atmospheric transport and Boltzmann equations for plasmas and neutron transport to other areas as chemical kinetics, population biology, firing rates in neuroscience, gradient flow transport in machine learning and optimisation, or the distribution of wealth in economics.
In all these cases the quantum structures discussed here apply.

\indent It is often believed that quantum mechanics is fundamentally different from classical statistics. 
An opposite view considers quantum systems as a particular case of classical probabilistic systems \cite{CWO2, CWPO, CWEQMCS}.
What singles out quantum mechanics from more general classical probabilistic systems is the unitary time evolution.
In classical statistics the expectation values of observables are computed from a (functional) integral over a probability distribution for events at all times and positions. 
For time-local observables this can be cast into the quantum rule in terms of classical wave functions and operators associated to observables.
The time evolution of the classical wave functions obeys a linear evolution equation.
For general classical probabilistic systems it is not unitary, however. 
The particular cases for which the evolution is unitary realise all features of a quantum system.

\indent For many simple quantum systems the associated classical probabilistic system has already been found \cite{CWPW2}. 
This includes probabilistic cellular automata \cite{CWFQFT, THO}, quantum field theories for fermions \cite{CWCWF}, or a small number of coupled qubits \cite{CWEQMCS}. 
The present note establishes that probabilistic transport equations are quantum systems. 
It supplements similar work for the Liouville equation \cite{CWQFTCF,CWDS,CWQO}, which can be considered as a special case of the systems discussed here. 

\indent Non-commuting operators appear in several circumstances in classical probabilistic systems. 
The transfer matrix does not commute with operators for classical observables, except for conserved quantities. 
For transport equations operator methods based on the pioneering work of Koopman \cite{KOOP} and von Neumann \cite{VNEU} have proven to be useful in many 
areas \cite{GOZ1, KAN, MAMA, MAU, GOMA, NIC, CHRU, NAKL, VOLO, BON, MEZI2, AME, KPM, KLE, KKB, NAME, MEZI1, MAUROY, MEZI, ACKA, DW, SGK, VAGI, NOJO, HAV}, an example being Liouville equations.
For related ideas see also ref.~\cite{GRT, CWET1, CWET2, BER}. 
In a certain sense, this operator approach corresponds to the Heisenberg picture for quantum mechanics.
We develop here the corresponding Schr\"odinger picture with wave functions encoding the time-local probabilistic information. 
Our approach differs from Koopman and von Neumann by considering a real classical wave function which is the square root of the time-local probability distribution.
A complex structure maps the real wave function to a complex wave function. 
In contrast to the Koopman--Von Neumann wave function the phases of the complex wave function matter for the time evolution according to a complex Schr\"odinger equation and for the expectation values of observables. 
This is important for the interference effects characteristic for quantum mechanics. 

\indent The operators for classical observables mutually commute.
We emphasize the important role of statistical observables.
They measure properties of the probabilistic information, conceptually similar to pressure or temperature.
The operators for statistical observables do not commute with the ones for classical observables and typically do not commute among themselves.
This realises the uncertainty principle characteristic for quantum mechanics.
For pairs of statistical observables, or statistical and classical observables, no classical correlation functions exists.
One may define a different correlation function \cite{CWO1, CWO2}, but this needs not to obey Bell's inequalities \cite{BEL1, CHSH}.
We do not see any characteristic quantum feature which cannot be realised in this classical probabilistic setting.

\section{Probabilistic transport equations}
\subsection*{Transport equations}

\indent We consider general first order deterministic evolution equations
\begin{equation}
\partial_t \sigma = F(\sigma)\,.
\label{eq:1}
\end{equation}
Here $\sigma(t)$ is taken as a vector with an arbitrary number of components, possibly infinitely many. 
For example, $\sigma(t,\vec x)$ may be a field in a $D$-dimensional space with coordinates $\vec x$.
The ``force'' $F$ is a function or functional of $\sigma(t)$. 
If $\sigma$ is a field and the evolution is local, the force typically involves space derivatives $\partial_{\vec{x}} \sigma(t,\vec x)$ at fixed $t$. 
It does not contain time-derivatives of $\sigma$, however.
The field $\sigma$ can be itself a probabilistic quantity, as for the Boltzmann equation.
We will often call a given value of the vector $\sigma$ a ``configuration'', since for the case where $\sigma$ is a field it is a given field configuration. 
We focus mainly on time-translation invariant equations for which $F$ does not depend explicitly on time.
If $F$ is even in $\sigma$ eq.~\eqref{eq:1} is time reversal invariant.
It retains its form if the sign of $t$ and $\sigma$ is switched.
If $\sigma(t)$ is a solution, also $-\sigma(-t)$ is a solution.
Time reversal symmetry can also be realised by a more general discrete transformation of $\sigma$
\begin{align}
T:\quad t\to &-t\,,\quad \sigma(t)\to A_T^{-1}\sigma(-t)\,,\quad A_T^2 = 1\,,
 \nn \\
&F(A_T\sigma) = -A_T F(\sigma).
\label{eq:1A}
\end{align}
Time reversal symmetry is, however, not necessary for our setting.
The quantum structure discussed here only requires that the map of the probabilistic information form $t$ to $t+\varepsilon$ is invertible.

\indent A prominent example for our setting is the Boltzmann equation where
$\sigma$ corresponds to the distribution function $f(\vec x,\vec v)$ in dependence on position $\vec x$ and velocity $\vec v$,
\begin{equation}
\sigma(t)=f(t;\vec x,\vec v)\,.
\label{eq:E1}
\end{equation}
The ``force'' reads
\begin{equation}
F(f)
=
-\vec v\,\partial_{\vec x}f
-\frac{\vec K}{m}\,\partial_{\vec v}f
+C[f]\,,
\label{eq:E2}
\end{equation}
where $\vec K$ is the force acting on single particles with mass $m$ and $C[f]$ is the collision term.
In the absence of $C$, and for $\vec K(\vec x)$ not dependent on
$f$, eq.~\eqref{eq:E2} reduces to the Liouville equation for the
probability distribution of particles as a function of their position and velocity. 
A probability distribution over the field $f(\vec x,\vec v)$ is a probability distribution over distribution functions.

\indent Important cases are the Vlasov equation for a collisionless plasma,
$C=0$, with $(\vec E,\vec B)$ electric and magnetic fields and $q$ the charge,
\begin{equation}
\vec K(\vec x,\vec v)
=
\frac{q}{m}
\left(
\vec E+\vec v\times\vec B
\right)\,,
\label{eq:E3}
\end{equation}
or the evolution of stars or galaxies under the influence of a gravitational potential $\phi(\vec x)$,
\begin{equation}
\vec K
=
-\partial_{\vec x}\phi(\vec x)\,.
\label{eq:E4}
\end{equation}
In the absence of the collision term and for $\vec K$ independent of $f$ the force $F[f]$ is linear in $f$.
In general, $\vec K$ can depend on $f$.
For example, the gravitational potential may be computed as a function of the distribution $f$ of particles.
In this case the evolution equation for $f$ is a non-linear differential equation.

Fluid dynamics obtains by integrating over $\vec v$,
$\rho(\vec x)=\int_{\vec v}f(\vec x,\vec v)$, $\rho(\vec x)\,\vec u(\vec x)=\int_{\vec v}\vec v\,f(\vec x,\vec v)$,
where for the mass we set $m=1$.
The field $\sigma$ may be taken now as
\begin{equation}
\sigma
=
\bigl(\rho(\vec x),\vec u(\vec x)\bigr)\,,
\label{eq:E5}
\end{equation}
with $F(\sigma)$ given by the continuity equation,
\begin{equation}
\partial_t\rho
=
-\partial_{\vec x}\bigl(\rho\vec u\bigr)\,,
\label{eq:E6}
\end{equation}
and the Navier--Stokes equation,
\begin{equation}
\partial_t\vec u
=
-\bigl(\vec u\,\partial_{\vec x}\bigr)\vec u
-\frac{1}{\rho}\partial_{\vec x}p
+\nu\,\partial_{\vec x}^{\,2}\vec u
+\vec K\,.
\label{eq:E7}
\end{equation}
Here $p$ is the pressure and $\nu=\mu/\rho$ is given by the viscosity $\mu$.
If one can derive an effective functional dependence
$\vec u(\rho)$, as for example for the flow of traffic, one has $\sigma = \rho (\vec x)$ with
(non-linear) force term given by eq.~\eqref{eq:E6}.

\indent Another important group of transport equations are reaction-diffusion equations for some concentration or scalar function $c(\vec x)$,
\begin{align}
&\sigma = c (\vec x) \,, \nn \\
&\partial_t c
=
-\partial_{\vec x}(c\vec u)
+
D\,\partial_{\vec x}^{\,2}c
+
R(c)\,,
\label{eq:E8}
\end{align}
where $\vec u(\vec x)$ is the given velocity of a fluid,
$D\,\partial_{\vec x}^{\,2}c$ the diffusion term and $R(c)$ the reaction term.
The concentration may have several components. This type of equation
describes many phenomena in meteorology, chemistry or biology. In the
absence of reactions the force term is linear in $\sigma$.
For $R=0$ this form of transport equation also describes the motion of a
particle under the influence of a stochastic force. If we take for $c(\vec x)$ the
probability to find the particle at $\vec x$ and identify $\vec u$ with
the drift, then it is the Fokker--Planck equation.

\indent For machine learning we can identify $\sigma$ with the data.
For example, one may learn a suitable ``force'' $F$ which transforms some probability distribution $w (t_{\text{in}}, \sigma)$ to some other probability distribution $w(t_f, \sigma)$ by a stepwise variable transformation of the data, $\sigma (t) \rightarrow \sigma(t+\varepsilon)$.
In this case $F$ is typically time-dependent.
\indent With probabilistic initial conditions for $\sigma$ our quantum formalism applies to all these transport equations.
We will later also add a stochastic force.
There are, of course, also much simpler transport equations for which $\sigma$ is a multi-component vector with finite dimension.
For example, the motion of a particle is described by $\sigma = \left(\vec x, \vec v\right)$.

\subsection*{Classical wave functions and operators}

\indent For a probabilistic setting the initial conditions are given at $t_{\rm in}$ by a probability distribution $w(t,\sigma)$.
The time evolution of the probability distribution follows the conservation equation
\begin{equation}
\partial_t w = - \partial_\sigma(Fw)\,.
\label{eq:1B}
\end{equation}
No further ingredients beyond this invertible transport equation are needed for establishing the quantum structure.
If $\sigma$ is a field, $w$ is a function of field configurations.
This includes the case where $\sigma$ itself is already a probability distribution.
In this case we consider probabilities for different probability distributions.

\indent We employ the classical wave function \cite{CWQP1, CWQP2, CWIT} $q(t,\sigma)$ which equals the root of $w$ up to a sign
\begin{equation}
w(t,\sigma)=q^2(t,\sigma)\,.
\label{eq:1C}
\end{equation}
In turn, the evolution of the wave function obeys
\begin{equation}
\partial_t q = - F \partial_\sigma q - \frac{1}{2}(\partial_\sigma F)\,q\,.
\label{eq:1D}
\end{equation}
The expectation value of an observable $A$ obtains from the wave function by the quantum rule
\begin{equation}
    \label{eq:5A}
    \langle A \rangle = q^T \hat{A} \,q = \int\! \tilde{\mathcal{D}}\sigma\, q(\sigma) \hat{A} \,q(\sigma)\,,
\end{equation}
where $\hat{A}$ is the operator associated to $A$.
The (functional) integral $\int \tilde{\mathcal{D}} \sigma$ integrates over all configurations at fixed $t$, with $\int \hat{\mathcal{D}}\sigma \,q^2 = \int \hat{\mathcal{D}}\sigma \,w =1$. 
For classical observables $A^{(\text{cl})}(\sigma)$ the operators are all diagonal, $\hat{A} (\sigma, \sigma^\prime) = A^{(\text{cl})}(\sigma)\, \delta(\sigma-\sigma^\prime)$.
With eq.~\eqref{eq:1C} the quantum rule follows from the classical statistical rule for expectation values.
All operators for classical observables commute.

\indent We can write eq.~\eqref{eq:1D} as a Schr\"odinger equation,
\begin{align}
\label{eq:1E}
&i\partial_t q = Hq\,,\\
&H = -i \left( F \partial_\sigma + \frac{1}{2} \left( \partial_\sigma F \right) \right) =  \frac{1}{2}\{F(\hat\sigma),\hat \gamma\}\,,
\nn
\end{align}
where $\{\hat A,\hat B\}=\hat A\hat B+\hat B\hat A$.
It involves the hermitian operators
\begin{equation}
\hat\sigma=\sigma\,,\quad
\hat \gamma = -i\frac{\partial}{\partial \sigma}\,,\quad
[\hat\sigma_k,\hat \gamma_l]=i\delta_{kl}\,.
\label{eq:1F}
\end{equation}
With $H^\dagger=H$ the evolution is unitary.
Since the evolution equation \eqref{eq:1D} or \eqref{eq:1E} is linear, the superposition principle for solutions follows.
The classical wave functions encode the wave aspect of the probabilistic information in quantum mechanics.
The superposition of two wave functions does not equal the addition of two probability distributions, leading to the interference characteristic for quantum mechanics.

\section{Statistical observables}

\indent We associate functions of the operators $\hat \gamma$ and $H$ to ``statistical observables'' \cite{CWPO, CWQFC, CWPW}. 
They do not take fixed values for the ``classical microstates'' or configurations $\sigma$, but rather measure properties of the probabilistic information. 
Their status is similar to temperature in classical statistical equilibrium. For example, $\gamma^2$ measures the ``roughness'' of the probabilistic information
\begin{equation}
\langle \gamma^2\rangle = q^T \hat \gamma^2 q
= \int \tilde{\mathcal{D}} \sigma \, (\partial_\sigma q)^2 \,.
\label{eq:1G}
\end{equation}
Since $q$, $i\hat \gamma$ and $iH$ are real quantities one has
\begin{equation}
\langle \gamma\rangle = 0\,,\quad \langle H\rangle = 0\,.
\label{eq:1H}
\end{equation}
(Strictly speaking, odd powers of $\gamma$ and $H$ do not act within the space of real
wave functions $q$ and may therefore not be considered as genuine observables.)
The eigenvalues of $H$ come in pairs with opposite signs.
The only possible eigenvalues for $H$ which are
compatible with real $q$ are the static states, $Hq=0$.

\indent Nevertheless, functions of $H$ and $\hat \gamma$ can be very
useful for understanding the dynamics of the system. 
For example, $H^2$ corresponds to a genuine statistical observable with associated operator acting in the space of real $q$.
With $\partial_k=\partial/\partial \sigma_k$, and summing over double indices one has
\begin{align}
H^2
&= \frac{1}{4}\Bigl(\{F(\hat\sigma),\hat \gamma\}^2\Bigr)
=
- \Bigl[
F_kF_l\,\partial_k\partial_l
+ (\partial_l(F_kF_l))\,\partial_k \nn \\
&+ \frac{1}{2}\partial_k(F_k\partial_l F_l)
- \frac{1}{4}(\partial_kF_k)^2
\Bigr]\,.
\label{eq:1I}
\end{align}
Consider the case where $F$ and therefore $H$ does not depend on time explicitly.
Then the squared quantum energy is a conserved quantity since its associated operator $H^2$ obviously commutes with $H$. 
The expectation value
\begin{equation}
\langle H^2\rangle
=
\int \tilde{\mathcal{D}} \sigma \,
\left(F_k\partial_k q + \frac{1}{2}(\partial_kF_k)q\right)^2
\label{eq:17}
\end{equation}
is independent of time. This is an important ingredient for stability in a situation where the eigenvalues of $H$ can take both positive and negative values. 
On the other hand, if $\langle H^2 \rangle$ differs from zero at initial time, this constitutes on obstruction for an approach to a static probability distribution.
Static wave functions are eigenfunctions of $H^2$ with eigenvalue zero.

\indent The spectrum of the quantum Hamiltonian $H$ can
be used for an investigation of probability distributions with periodic time evolution. 
For the non-zero eigenvalues of $H$ the eigenfunctions $\bar \varphi$ are not real
\begin{equation}
\varphi_n(t)=\bar\varphi_n(\sigma)\exp(-iE_nt)\,,\quad
H\bar\varphi_n(\sigma)=E_n\bar\varphi_n(\sigma)\,.
\label{eq:1L}
\end{equation}
Real oscillatory wave functions can be constructed by taking the real part of $\varphi_n(t)$,
$\bar\varphi_n(\sigma)=|\bar\varphi_n(\sigma)|\exp(i\alpha_n(\sigma))$,
\begin{equation}
q_n(t)=\Re(\varphi_n(t))
=
|\bar\varphi_n(\sigma)|\cos\bigl(\alpha_n(\sigma)-E_nt\bigr)\,.
\label{eq:1M}
\end{equation}
They are eigenfunctions of the operator $H^2$,
\begin{equation}
H^2 q_n(t)=E_n^2 q_n(t)\,.
\label{eq:1N}
\end{equation}
The oscillating wave function \eqref{eq:1M} yields directly a periodic probability distribution by taking the square \eqref{eq:1C}.
The quantum formalism establishes the existence of periodic probability distributions within the very general setting of transport equations. 
We emphasize that the quantum energy $E_n$ differs from the classical energy $E^{(\text{cl})}$ of the system.
The eigenvalues of $H$ describe the frequencies of periodic probability distributions.

\indent As familiar in quantum mechanics we can introduce operators for the time-derivatives of observables
\begin{equation}
\partial_t \langle A\rangle = q^T \hat{\dot A} \,q\,, \quad \hat{\dot A} = i[H, \hat A]\,.
\label{eq:15A2}
\end{equation}
In particular, one has
\begin{align}
\hat{\dot \sigma} &= i[H,\hat\sigma] = F(\hat\sigma)\,, \nonumber\\
\hat{\dot \gamma}_k &= i[H,\hat \gamma_k]
= -(\partial_k F_l)(\hat \sigma)\hat \gamma_l
+ \frac{i}{2}\,\partial_k\partial_l F_l(\hat\sigma)\,.
\label{eq:1K}
\end{align}
This permits us to compute the time derivative of the roughness $\gamma^2$ from the time-local probabilistic information contained in the wave function $q(t)$.
If $\hat{A}$ commutes with $H$ the observable $A$ is a conserved quantity.
This includes statistical observables.

\indent The transport equation may be invariant under some continuous symmetry transformation acting on $\sigma$, with infinitesimal $\alpha_i$,
\begin{equation}
    \label{eq:15A}
    \delta \sigma_j = \alpha_i \left( B_{ijk}\sigma_k + C_{ij} \right)\,.
\end{equation}
This translates to the transformation of the wave function
\begin{align}
    \delta q &= \frac{\partial q}{\partial \sigma_j} \delta \sigma_j = \alpha_i \left( B_{ijk}\sigma_k\frac{\partial}{\partial \sigma_j} + C_{ij}\frac{\partial}{\partial \sigma_j}  \right)q \nn \\
    &=i \alpha_i \hat{L}_i q\,.
    \label{eq:15B}
\end{align}
The generators of symmetry transformations commute with $H$,
\begin{equation}
    \label{eq:15C}
    \left[ \hat{L}_i, H \right]=0\,, \quad \hat{L}_i^\dagger = \hat{L}_i\,.
\end{equation}
They define conserved statistical observables.
For example, for a symmetry $SO(3)$ of rotations of components of $\sigma$ these are the components of generalized quantum angular momentum,
\begin{equation}
    \label{eq:15D}
    (\hat{L}_i )_{jk} = -i \varepsilon_{ijk} \,\sigma_j \frac{\partial}{\partial \sigma_k}\,.
\end{equation}
For non-abelian symmetries the generators $\hat L_i$ do not commute.
If $F$ contains only derivatives of $\sigma$ as for membranes, the shift symmetry $\sigma \to \sigma + \alpha$ induces a (generalized) conserved momentum observable.
It coincides with the roughness $\gamma$ in eq.~\eqref{eq:1E}.

\indent If $\sigma$ is a field, important symmetries are translations or rotations in space. 
The associated conserved statistical observables are the quantum momentum and angular momentum. 
These observables differ from the classical momentum and angular momentum. 
Even functions of symmetry generators constitute an important class of statistical observables. 
The non-commuting properties of their associated operators follow directly from a non-commuting structure of the symmetry transformations.

\section{Complex wave function}

\indent As an important advantage of the use of wave functions one can perform basis transformations. 
We employ a Fourier transform in order to introduce a complex structure for the wave function,
\begin{equation}
\psi(t,\gamma)=\int \tilde{\mathcal{D}} \sigma \,\exp(-i\gamma\sigma)\,q(t,\sigma)\,.
\label{eq:1O}
\end{equation}
The wave function $\psi$ is complex, obeying the
constraint
\begin{equation}
\psi(t,-\gamma)=\psi^\ast(t,\gamma)\,.
\label{eq:1P}
\end{equation}
Incorporating the $2\pi$-factors of Fourier transforms in $\int \tilde D\gamma$ the wave function is normalised,
\begin{equation}
    \label{eq:1PA}
    \psi^\dagger \psi = \int \tilde{\mathcal{D}}\gamma \,\psi^\ast(t,\gamma)\psi(t,\gamma)=1\,.
\end{equation}
In the Fourier basis the operators are given by
\begin{align}
&\hat\sigma = i\frac{\partial}{\partial \gamma}\,,\quad
\hat\gamma=\gamma\,, \nn \\
&H=\frac{1}{2}\left\{F\!\left(i\frac{\partial}{\partial \gamma}\right),\gamma\right\}.
\end{align}
(In some cases, as for the Liouville equation, it may be advantageous to perform the Fourier transform only for part of the variables \cite{CWQFTCF}.)

\indent The phases of $\psi$ are important for the time evolution according to the complex Schr\"odinger equation
\begin{equation}
i\partial_t \psi(t,\gamma)=H\psi(t,\gamma)\,,
\label{eq:1Q}
\end{equation}
and for superpositions of solutions which contain the interference effects characteristic for quantum mechanics. 
Different phases correspond to different real wave functions $q(\sigma)$,
\begin{align}
\psi(\gamma)&=|\psi(\gamma)|\exp(i\delta(\gamma))\,, \nn \\
q(\sigma)
&=
\int \tilde{\mathcal{D}}\gamma \,\exp(i\gamma\sigma)\,\psi(\gamma)
\nonumber\\
&=
\int \tilde{\mathcal{D}}\gamma \,|\psi(\gamma)|\cos\bigl(\sigma\gamma+\delta(\gamma)\bigr)\,.
\label{eq:1R}
\end{align}
The constraint \eqref{eq:15A} yields a relation between phases $\delta(\gamma)$ and amplitudes $|\psi(\gamma)|$ for positive and negative $\gamma$.
It may be advantageous to solve the Schr\"odinger equation \eqref{eq:15D} for a general wave function $\tilde \psi (t,\gamma)$ without taking into account the constraint \eqref{eq:15A}.
One may then define 
\begin{equation}
    \label{eq:21A}
    \psi(t,\gamma) = (\tilde{\psi}(t,\gamma) + \tilde{\psi}^\ast(t,-\gamma))/2 = \frac{1}{2}(1+C)\tilde \psi\,.
\end{equation}
With $[C,H]=0$ this constitutes a solution of the Schr\"odinger equation which respects the constraint.
The operation of $C$ involves a complex conjugation, similar to charge conjugation.
One finds similarly to eq.~\eqref{eq:1R}
\begin{align}
    &\tilde{\psi} (\gamma) = |\tilde{\psi} (\gamma)| e^{i\tilde{\delta}(\gamma)}\,, \nn \\
    &q(\sigma) = \int \!\hat{\mathcal{D}}\gamma \, |\tilde{\psi} (\gamma)| \cos( \sigma\gamma + \tilde{\delta}(\gamma) )\,.
    \label{eq:21B}
\end{align} 

\indent At this point it seems rather obvious that the probabilistic
setting for the classical transport equation \eqref{eq:1} constitutes a quantum system. 
The unitary evolution of the complex wave function follows the Schr\"odinger equation with hermitian Hamiltonian $H$. 
The expectation values of observables are computed from the quantum rule using appropriate operators,
\begin{equation}
\langle A(t)\rangle = \int \tilde{\mathcal{D}}\gamma \,\psi^\ast(t,\gamma)\,\hat A\,\psi(t,\gamma)\,.
\label{eq:1S}
\end{equation}
This includes statistical observables. 
The operators for statistical observables typically do not commute with the
ones for the classical observables. 
This induces Heisenberg's uncertainty principle for pairs of observables represented by non-commuting operators.
One can define the usual quantum correlations for such pairs of observables. 
They do not have to obey Bell's inequalities for classical correlation functions. 
Classical correlation functions simply do not exist for statistical observables, since those have no definite values in the microstates of the system. 
The only particularity is the form of the Hamiltonian. 
The quantum energy associated to $H$ differs from the classical energy. 
It is a statistical observable related to the time-periodicity of the probability distribution. 
Periodic probability distributions correspond to sharp values of the observable $H^2$.
A situation even closer to familiar quantum systems arises if one can find closed subsystems for which the Hamiltonian is bound \cite{CWQFTCF}.

\section{Discrete time evolution}

\indent Probabilistic classical systems described by a transport equation \eqref{eq:1} can be investigated numerically. 
This concerns both the time evolution of the wave function or associated probability distribution, and a functional integral approach that we will discuss later. 
Numerical solutions typically proceed by a discretization in time and, if appropriate, in space. 
This discretization should respect the linearity and unitarity of the evolution law.
It provides a regularisation for which all mathematical expressions are well defined.

\subsection*{Probabilistic automata}

\indent We describe the corresponding discrete systems as probabilistic automata \cite{CWFQFT}.
For $\sigma$ playing the role of local fields these are cellular automata \cite{ULA, VNEUA, ZUS, WOL, THO}.
The updating of the automaton is deterministic, representing the transport equation \eqref{eq:1}. 
Initial conditions are given by a probability distribution or wave function, which induces the probabilistic properties crucial for quantum physics. 
We can also consider the discrete systems as genuine dynamical systems. 
Our quantum picture therefore describes probabilistic invertible automata as quantum systems.

\indent A key point for the discrete systems discussed here
is the unitary time evolution of the wave function.
Instead of the Schr\"odinger equation with hermitian Hamiltonian
the evolution of the wave function
in discrete time steps $\varepsilon$ is governed by a unitary
step evolution operator. If appropriate, a Schr\"odinger
equation obtains from there by the continuum limit
$\varepsilon\to0$. There are two possible approaches. The first
treats $\sigma$ as discrete variables. A set of discrete
variables $\sigma_i(t)$ is updated by an invertible map
to a new set $\sigma_i(t+\varepsilon)$. For each configuration
$\sigma_i(t)$ there exists precisely one configuration
$\sigma_i(t+\varepsilon)$, and so on for larger $t$. This defines
a trajectory of configurations $\sigma_i(t')$
which are connected to $\sigma_i(t)$ by
the updating. The value of the wave function is
the same for all configurations on the trajectory. The
difference $\sigma_i(t)-\sigma_j(t)$ between configurations
belonging to two different trajectories $i$ and $j$
may depend on $t$. Nevertheless, the normalisation
of the wave function for all $t$ is guaranteed by an
invertible updating. A unitary time evolution
is realised directly. At every time the integration
$\int \tilde{\mathcal{D}}\sigma(t)$ is specified by the sum over the discrete
trajectories.

\indent The second approach uses a fixed integration measure
for $\int \tilde{\mathcal{D}}\sigma(t)$ for all $t$. For example, $\sigma_k(t)$ may
be real numbers and $\int \tilde{\mathcal{D}}\sigma(t)=\prod_k \int_{-\infty}^{\infty} d\sigma_k(t)$.
In this case the position of the trajectories $\sigma_i(t)$ in the space of real numbers matters. 
A unitary evolution needs to take this into account, which corresponds
to an explicit implementation of the term $-(i/2)\partial_\sigma F$ in the continuum version \eqref{eq:1E}.
This term is always generated in the limit of continuous $\sigma$. 
For the first approach it results from the way how the limit from a sum over discrete trajectories to a continuum integral is taken. 
For the second approach no particular continuum limit for the integration $\int \tilde{\mathcal{D}}\sigma(t)$ has to be taken. 
The second approach is more suitable for the formulation of a functional integral. 
We will start with the first approach which emphasizes the properties of probabilistic automata, and turn to the second approach subsequently.

\indent We consider a deterministic evolution with discrete time steps $\varepsilon$,
\begin{equation}
\sigma(t+\varepsilon)=\sigma(t)+\varepsilon F(t+\varepsilon)\,,
\label{eq:2}
\end{equation}
where $F(t+\varepsilon)$ is a function (or functional) of the configurations $\sigma(t+\varepsilon)$. 
Eq.~\eqref{eq:2} encodes the updating rule for an automaton. 
We require an invertible updating. 
For any configuration $\sigma_i(t+\varepsilon)$ there exists a unique configuration $\sigma_i(t)$ from which it originates by the updating. 
This should hold for an arbitrary set of discrete configurations $\sigma_i(t+\varepsilon)$ labelled by $i$. 
For the argument of $F$ we could express $\sigma(t+\varepsilon)$ as a function of $\sigma(t)$. 
The particular form \eqref{eq:2} is most convenient for the formulation of the step evolution operator. 
If eq.~\eqref{eq:2} is considered as a
discretization of the continuous transport equation \eqref{eq:1},
the latter should be reproduced in the limit $\varepsilon\to0$,
with $\partial_t\sigma=(\sigma(t+\varepsilon)-\sigma(t))/\varepsilon$. 
This requirement leaves a certain freedom in the choice of $F(t+\varepsilon)$ for finite $\varepsilon$. 
One may exploit this in order to implement time reversal symmetry in the
discrete formulation if the continuous system is time reversal invariant according to eq.~\eqref{eq:1A}.
The time-reversed evolution step switches the sign of $\varepsilon$
\begin{equation}
\sigma(t-\varepsilon)=\sigma(t)-\varepsilon F(\sigma(t-\varepsilon))\,.
\label{eq:3}
\end{equation}
Time reversal symmetry is realised in the discrete formulation if one can replace in eq.~\eqref{eq:3} $F(\sigma(t-\varepsilon))$ by $F(\sigma(t))$. 
For a time-reversal invariant continuum equation this can be achieved by a suitable symmetrization of $F$, for example by alternating the updating \eqref{eq:2} with the time reversed one \eqref{eq:3} for consecutive steps.

\indent At some initial time $t_{\rm in}$ one specifies probabilistic initial conditions for the discrete evolution equation \eqref{eq:2}.
As before we employ the classical wave function $q(t)$ which is a real function of the configurations $\sigma(t)$, with the time-local probability for a given configuration $\sigma(t)$ given by $w(t,\sigma(t))=q^2(t,\sigma(t))$.
The wave function is a unit vector,
\begin{equation}
\int \tilde{\mathcal{D}}\sigma(t)\, q^2\bigl(t,\sigma(t)\bigr)=1\,,
\label{eq:5}
\end{equation}
such that the positivity and normalisation of the time-local probability distribution is guaranteed.
If $\sigma$ is a field, the integral $\int \tilde{\mathcal{D}}\sigma(t)$ is a
functional integral over functions $\sigma(t,\vec x)$ at a given $t$.
For a discrete set of configurations $\sigma_i$ the integral $\int \tilde{\mathcal{D}}\sigma(t)$ amounts simply to a sum over $i$. 
The probabilistic initial condition specifies the wave function $q(t_{\rm in})$.

\subsection*{Step evolution operator}

\indent The time evolution of the wave function is
governed by an orthogonal step evolution operator $\hat S$,
\begin{align}
q(t+\varepsilon)&=\hat S(t)\,q(t)\,, \\
q\bigl(t+\varepsilon,\sigma(t+\varepsilon)\bigr)
&=
\int \tilde{\mathcal{D}}\sigma(t)\,
\hat S\bigl(t;\sigma(t+\varepsilon),\sigma(t)\bigr)\,
q\bigl(t,\sigma(t)\bigr) \nn
\label{eq:6}
\end{align}
The precise notion of orthogonality depends on the measure $\int \tilde{\mathcal{D}}\sigma(t)$. 
We first consider the configurations $\sigma(t)$ as points on a discrete set of trajectories $i$, with $\int \tilde{\mathcal{D}}\sigma(t) \equiv \sum_i$.
In this case the step evolution is given by
\begin{equation}
\hat S\bigl(t;\sigma(t+\varepsilon),\sigma(t)\bigr)
=
\delta\bigl(\sigma(t+\varepsilon)-\sigma(t)-\varepsilon F(t+\varepsilon)\bigr)\,,
\label{eq:7}
\end{equation}
with
\begin{equation}
q\bigl(t+\varepsilon,\sigma(t+\varepsilon)\bigr)
=
q\bigl(t,\sigma(t+\varepsilon)-\varepsilon F(t+\varepsilon)\bigr)\,.
\label{eq:7A}
\end{equation}
The $\delta$-function in eq.~\eqref{eq:7} involves discrete Kronecker symbols. 
For a finite number of trajectories $\hat S(t)$ is a finite-dimensional unique jump matrix with one entry equal to one in each row and column, and zero entries otherwise.
Such a matrix is indeed orthogonal. The normalisation of the wave function is preserved accordingly.
The matrix $\hat S$ is positive semidefinite in the  sense that all  its elements are positive or zero. 
Eq.~\eqref{eq:7A} enforces that at $t+\varepsilon$ the value of the wave function for the configuration $\sigma(t+\varepsilon)$ is the same as the one for the wave function at $t$ for the configuration $\sigma(t)$ from which $\sigma(t+\varepsilon)$ originates by the updating rule \eqref{eq:2}. 
The same holds for the time-local probabilities. 
As it should be for a deterministic evolution, the probabilities at $t$ are transported to $t+\varepsilon$ by the updating.
The step evolution operator $\hat S$, together with the probabilistic initial wave function, describes a probabilistic automaton with deterministic updating and probabilistic initial condition. 
For fields with an evolution equation which is local in space the formulation with discrete positions $\vec{x}$ deals with a cellular automaton.
The cells are labeled by the positions $\vec{x}$.

\indent We next switch to a fixed integration measure $\int \tilde{\mathcal{D}}\sigma(t)$. 
In this case one needs a rescaling of $q(t+\varepsilon)$ after the updating \eqref{eq:7A}, such that the normalisation \eqref{eq:5} is preserved in the step from $t$ to $t+\varepsilon$
\begin{align}
\label{eq:7B}
q\bigl(t+\varepsilon,&\sigma(t+\varepsilon)\bigr)= \\
&\bigl(1+\varepsilon \mathcal{N}(\sigma(t+\varepsilon))\bigr)\,
q\bigl(t,\sigma(t+\varepsilon)-\varepsilon F(t+\varepsilon)\bigr)\,. \nn
\end{align}
Correspondingly, the step evolution operator becomes in this setting
\begin{equation}
\hat S(t)
=
\bigl(1+\varepsilon \mathcal{N}(\sigma(t+\varepsilon))\bigr)\,
\delta\bigl(\sigma(t+\varepsilon)-\sigma(t)-\varepsilon F(\sigma(t+\varepsilon))\bigr).
\label{eq:7F}
\end{equation}
The factor $(1+\varepsilon \mathcal{N})^2$ is given by the Jacobian of the variable transformation from
$\sigma-\varepsilon F(\sigma)$ to $\sigma$,
\begin{equation}
\bigl(1+\varepsilon \mathcal{N}(\sigma)\bigr)^2
=
\det\!\left(
\delta_{kl}-\varepsilon \frac{\partial F_k}{\partial \sigma_l}
\right)\,.
\label{eq:7E}
\end{equation}
In the continuum limit eq.~\eqref{eq:7B} yields
\begin{equation}
\partial_t q(t)
=
-\,F\,\frac{\partial}{\partial\sigma}\,q(t)
+
\mathcal{N}\,q(t)\,,
\label{eq:7C}
\end{equation}
and we recognize eq.~\eqref{eq:1D},
\begin{equation}
\mathcal{N}
=
-\frac12\,\frac{\partial F}{\partial \sigma}
=
-\frac12 \sum_k \partial_k F_k(\sigma)\,.
\label{eq:7D}
\end{equation}
Eq.~\eqref{eq:7E} yields indeed eq.~\eqref{eq:7D} in the limit $\varepsilon\to0$. 

\indent We can perform the Fourier transform \eqref{eq:1O} for the step evolution operator
\begin{align}
\hat S_{\gamma}(t)
&=\int \tilde D \sigma(t+\varepsilon)\,\tilde D \sigma(t)
\exp\bigl(-i\gamma(t+\varepsilon)\sigma(t+\varepsilon)\bigr) \hat S(t)
\nn \\
&\quad
\times\exp\bigl(i\gamma(t)\sigma(t)\bigr)
\nonumber\\
&=
\int \tilde D \sigma(t+\varepsilon)
\exp\Bigl(
-i\Bigl[\sigma(t+\varepsilon)\bigl(\gamma(t+\varepsilon)-\gamma(t)\bigr)
\nn \\
&\qquad \qquad
+\varepsilon F(\sigma (t+\varepsilon))\gamma(t)\Bigr]
\Bigr)\,
\bigl(1+\varepsilon \mathcal{N}(\sigma(t+\varepsilon))\bigr)\,.
\label{eq:7G}
\end{align}
It governs the evolution of the complex wave function,
\begin{align}
\label{eq:7H}
&\psi(t+\varepsilon)=\hat S_{\gamma}(t)\,\psi(t)\,, \\
&\psi\bigl(t+\varepsilon,\gamma(t+\varepsilon)\bigr)
= \nn\\
&\hspace{1cm}\int \tilde D \gamma(t)\,
\hat S_{\gamma}\bigl(t;\gamma(t+\varepsilon),\gamma(t)\bigr)\,
\psi\bigl(t,\gamma(t)\bigr). \nn
\end{align}
Here the integral $\int \tilde D \gamma(t)$ incorporates the appropriate
$1/(2\pi)$-factors appropriate for the inverse Fourier transform,
\begin{equation}
\int \tilde D \gamma(t)\,
\exp \bigl\{i\gamma(t)\bigl(\sigma(t)-\sigma'(t)\bigr)\bigr\}
=
\delta \bigl(\sigma(t)-\sigma'(t)\bigr)\,.
\label{eq:7I}
\end{equation}
The normalisation of the complex wave function is preserved by the evolution \eqref{eq:7H} for all $t$,
\begin{equation}
\int \tilde{\mathcal{D}} \gamma(t)\,
\psi^*\bigl(t,\gamma(t)\bigr)\psi\bigl(t,\gamma(t)\bigr)
=
\int \tilde{\mathcal{D}} \sigma(t)\,q^2\bigl(t,\sigma(t)\bigr)=1\,.
\label{eq:7J}
\end{equation}
The step evolution operator $\hat S_{\gamma}$ is a unitary matrix.
The continuum limit $\varepsilon\to0$ of eq.~\eqref{eq:7H} yields the Schr\"odinger
equation \eqref{eq:1Q} for the complex wave function $\psi$.

\section{Functional integrals}

\indent The concepts of wave functions and operators are adapted to a time-local description of physical phenomena.
On a fundamental level, and also often for practical purposes, a global view of the probabilistic setting is appropriate.
It is based on probabilities for possible events at all times.
The time-local description obtains from the overall probability distribution by ``integrating out'' the past and the future \cite{CWPW}.
For probabilistic automata the relation between the time-local description and the overall probability distribution is rather simple.
Only the trajectories allowed by the automaton contribute with non-zero probabilities.

\subsection*{Euclidean functional integral}

\indent The overall probability distribution $\bar w[\sigma]$ is a functional
of $\sigma(t)$ for all times. It is defined by
\begin{equation}
\bar w[\sigma]
=
q\bigl(t_f,\sigma(t_f)\bigr)\,
\exp\bigl(-S[\sigma]\bigr)\,
q\bigl(t_{\mathrm{in}},\sigma(t_{\mathrm{in}})\bigr)\,,
\label{eq:8}
\end{equation}
with Euclidean action given by a product of step evolution operators
\begin{equation}
\exp\bigl(-S[\sigma]\bigr)
=
\prod_{t=t_{\mathrm{in}}}^{t_f-\varepsilon}
\hat S\bigl(t;\sigma(t+\varepsilon),\sigma(t)\bigr)\,.
\label{eq:9}
\end{equation}
For a positive semi-definite step evolution operator $\hat S$ the product on
the r.h.s. is positive or zero for all overall configurations. The action $S$ is therefore
real. It diverges if the r.h.s vanishes.
The overall probability differs from zero only for the trajectories allowed
by the updating rule \eqref{eq:2}. Each initial configuration $\sigma(t_{\mathrm{in}})$
defines a trajectory of configurations $\sigma(t)$ for all $t$.
They obtain from $\sigma(t_{\mathrm{in}})$ by the deterministic updating. For any overall
configuration which does not belong to one of the trajectories $\bar w[\sigma]$
vanishes due to the $\delta$-functions in $\hat S$. 
The probability for a particular trajectory labeled by
$\sigma(t_{\mathrm{in}})$ is given by $q^2\bigl(t_{\mathrm{in}},\sigma(t_{\mathrm{in}})\bigr)\ge 0$.
Indeed, $\bar w$ differs from zero only if $\sigma(t_f)$ belongs to the
trajectory originating from $\sigma(t_{\mathrm{in}})$. Since the value of the
wave function remains the same for all configurations on the trajectory,
$q\bigl(t_f,\sigma(t_f)\bigr)$ equals
$q\bigl(t_{\mathrm{in}},\sigma(t_{\mathrm{in}})\bigr)$ for all pairs
$\bigl(\sigma(t_f),\sigma(t_{\mathrm{in}})\bigr)$ belonging to one given trajectory. 
This shows that $\bar w$ is positive for arbitrary signs of the components of $q\bigl(t_{\mathrm{in}},\sigma(t_{\mathrm{in}})\bigr)$.
The normalisation of $\bar w$ follows from
\begin{equation}
Z
=
\int \mathcal D \sigma\,\bar w[\sigma]
=
\int \tilde{\mathcal{D}} \sigma(t_{\mathrm{in}})\,
q^2\bigl(t_{\mathrm{in}},\sigma(t_{\mathrm{in}})\bigr)
=
1\,.
\label{eq:10}
\end{equation}
Here $\int \mathcal D \sigma$ integrates over configurations at all $t$,
while $\int \tilde{\mathcal{D}} \sigma(t)$ integrates over the configurations at fixed $t$.

\indent Classical observables $A[\sigma]$ are functionals of the configurations at
all $t$. 
(This includes, for example, correlations as $\sigma (t_1, \vec{x})\, \sigma (t_2, \vec{y})$ for $t_2 \neq t_1$.)
Their expectation value is given by the classical statistical rule
\begin{equation}
\langle A\rangle
=
\int \mathcal D \sigma\,A[\sigma]\,\bar w[\sigma]\,.
\label{eq:11}
\end{equation}
Time-local classical observables depend only on configurations $\sigma(t)$ at a given $t$.
With
\begin{align}
    q(t, \sigma(t)) &= \prod_{t^\prime=t_{\mathrm{in}}}^{t-\varepsilon} \hat{S}(t^\prime)\, q(t_{\mathrm{in}}, \sigma(t_{\mathrm{in}})) \nn \\
    &= q(t_f, \sigma(t_f)) \prod_{t^\prime = t}^{t_f-\varepsilon} \hat{S}(t^\prime)\,,
    \label{eq:41A}
\end{align}
one infers from eqs.~\eqref{eq:11}, \eqref{eq:9}
\begin{equation}
\langle A(t)\rangle
=
\int \tilde{\mathcal{D}} \sigma(t)\,
q\bigl(t,\sigma(t)\bigr)\,
A\bigl(\sigma(t)\bigr)\,
q\bigl(t,\sigma(t)\bigr)\,.
\label{eq:12}
\end{equation}
This is the quantum rule
\begin{equation}
\langle A(t)\rangle
=
q^T\,\hat A\,q\,,
\label{eq:13}
\end{equation}
with a diagonal operator $\hat A$ associated to $A(t)$,
\begin{equation}
\hat A\bigl(\sigma(t),\sigma^\prime(t)\bigr)
=
A\bigl(\sigma(t)\bigr)\,
\delta\bigl(\sigma(t)-\sigma^\prime(t)\bigr)\,.
\label{eq:14}
\end{equation}

\indent The expectation values of statistical observables can be evaluated from a
functional integral by inserting the associated operator at suitable positions
in the chain of step evolution operators \eqref{eq:9}, and performing the functional
integral $\int \mathcal D \sigma$. The position of the insertion and the order
of the operators matter. This differs from the insertion of operators for
classical observables.

\subsection*{Extended overall weight distribution}

\indent One may be interested in the expectation values of observables whose associated operators involve both $\hat \sigma$ and $\hat \gamma$.
For this purpose we define an extended overall weight distribution which depends on both $\sigma$ and $\gamma$.
This will make contact to the MSRJD functional integral \cite{MSR, JAN, DOM}, for which $\gamma$ plays the role of the response field.
The extended overall weight dsitribution is defined by 
\begin{align}
\tilde w[\sigma,\gamma]
=
&\varphi^*\bigl(t_f,\sigma(t_f+\varepsilon),\gamma(t_f)\bigr)\,
\exp\bigl(iS_M[\sigma,\gamma]\bigr) \nn \\
&\times
\varphi\bigl(t_{\mathrm{in}},\sigma(t_{\mathrm{in}}),\gamma(t_{\mathrm{in}})\bigr)\,,
\label{eq:21}
\end{align}
with
\begin{align}
S_M
=
-\sum_{t=t_{\mathrm{in}}}^{t_f-\varepsilon}
\Big[
&\sigma(t+\varepsilon)\bigl(\gamma(t+\varepsilon)-\gamma(t)\bigr)
+\varepsilon F(t+\varepsilon)\gamma(t)
\nn \\
&+\varepsilon\,\Delta\bigl(\sigma(t+\varepsilon)\bigr)
\Big]\,,
\label{eq:22} \\
&\hspace{-1.5cm}\Delta\bigl(\sigma(t+\varepsilon)\bigr)
=
\frac{i}{2\varepsilon}\,
\mathrm{tr}\,
\ln\!\left(
\delta_{kl}
-
\varepsilon\frac{\partial F_k}{\partial \sigma_l} (t+\varepsilon)
\right)\,.
\nn
\end{align}
The boundary wave functions are given by using the extended wave function,
\begin{equation}
\varphi\bigl(t,\sigma(t),\gamma(t)\bigr)
=
\exp\bigl(-i\sigma(t)\gamma(t)\bigr)\,
q\bigl(t,\sigma(t)\bigr)\,,
\label{eq:23}
\end{equation}
at $t_{\mathrm{in}}$ and $t_f$, respectively. 

\indent The overall probability distribution $\bar w[\sigma]$ obtains from $\tilde w[\sigma,\gamma]$
by functional integration over $\gamma$,
\begin{equation}
\bar w[\sigma]
=
\int \mathcal D \gamma\,
\tilde w[\sigma,\gamma]\,.
\label{eq:24}
\end{equation}
On the other hand, the integration over $\sigma$ yields
\begin{equation}
\hat w[\gamma]
=
\int \mathcal D \sigma\,
\tilde w[\sigma,\gamma]\,,
\label{eq:25}
\end{equation}
where
\begin{equation}
\hat w[\gamma]
=
\psi^*\bigl(t_f,\gamma(t_f)\bigr)\,
\prod_{t=t_{\mathrm{in}}}^{t_f-\varepsilon}
\hat S_{\gamma}(t)\,
\psi\bigl(t_{\mathrm{in}},\gamma(t_{\mathrm{in}})\bigr)\,.
\label{eq:26}
\end{equation}
The normalisation is given by
\begin{equation}
Z
=
\int \mathcal D \sigma\,\mathcal D \gamma\,\tilde w[\sigma,\gamma]
=
\int \mathcal D \sigma\,\bar w[\sigma]
=
\int \mathcal D \gamma\,\tilde w[\gamma]
=
1\,.
\label{eq:27}
\end{equation}
The extended weight distribution $\tilde w$ contains information both on $\bar w [\sigma]$ and $\hat w [\gamma]$.
The functional $\hat w[\gamma]$ is convenient for a simple computation of observables for which the associated operators involve $\partial / \partial \hat \sigma$.
We may consider $\hat w [\gamma]$ as the (functional) Fourier transform of $\bar w [\sigma]$.
It is a complex quantity and therefore no longer a probability distribution.
Nevertheless, it demonstrates how complex weight distributions can encode the information of a real positive probability distribution.

\indent The extended wave function evolves according to the extended step evolution operator
\begin{align}
\varphi(t+\varepsilon)=\bar S(t)\varphi(t)\,,
\quad
\bar S(t)=\exp\bigl(-i\varepsilon \bar H(t)\bigr)\,,
\nn \\
\bar H(t)
=
\sigma(t+\varepsilon)\,\partial_t\gamma(t)
+
F(t+\varepsilon)\,\gamma(t)
+
\Delta\bigl(\sigma(t+\varepsilon)\bigr)\,,
\label{eq:28}
\end{align}
where we employ the discrete time derivative
\begin{equation}
\partial_t\gamma(t)
=
\frac{1}{\varepsilon}\bigl(\gamma(t+\varepsilon)-\gamma(t)\bigr)\,,
\label{eq:29}
\end{equation}
and
\begin{equation}
\bar S(t)\varphi(t)
=
\int \tilde{\mathcal{D}} \sigma(t)\,\tilde{\mathcal{D}} \gamma(t)\,
\exp\bigl(-i\varepsilon \bar H(t)\bigr)\,
\varphi(t)\,.
\label{eq:30}
\end{equation}
The evolution according to eq.~\eqref{eq:28} preserves the relation \eqref{eq:23} between $\varphi$ and $q$ for all $t$, as one may verify with eqs.~\eqref{eq:7B}, \eqref{eq:7E}.
These relations can be seen by writing in eq.~\eqref{eq:21}
\begin{equation}
\exp(iS_M)
=
\prod_{t=t_{\mathrm{in}}}^{t_f-\varepsilon}
\bar S(t)\,,
\label{eq:63A}
\end{equation}
and noting the relation between $\bar S(t)$ and $\hat S(t)$,
\begin{align}
\bar S(t)=\int \tilde{\mathcal D}\sigma(t)
\exp\{
&-i[\gamma(t+\varepsilon)\sigma(t+\varepsilon)
\nn \\
&-\gamma(t)\sigma(t)]\}\hat S(t)\,.
\label{eq:63B}
\end{align}

\indent The continuum limit of the action $S_M$ in eq.~\eqref{eq:22} reads
\begin{align}
S_M
=
-\int_t
\Bigg\{
&\sigma(t)\,\partial_t\gamma(t)
+
F\bigl(\sigma(t)\bigr)\,\gamma(t)
\nn \\
&-\frac{i}{2}\sum_k
\frac{\partial F_k}{\partial \sigma_k}\bigl(\sigma(t)\bigr)
\Bigg\}\,.
\label{eq:32}
\end{align}
By partial integration one can replace
$\sigma\,\partial_t\gamma \to -\gamma\,\partial_t\sigma$,
\begin{equation}
e^{iS_M}
=
\exp\!\left\{
\int_t
\left[
i\gamma(t)\bigl(\partial_t\sigma(t)-F(t)\bigr)
-\frac{1}{2}\frac{\partial F}{\partial \sigma}(t)
\right]
\right\}\,.
\label{eq:33}
\end{equation}
This corresponds to the functional integral of ref.~\cite{MSR, JAN, DOM} once the contribution from a stochastic force is omitted. 
The last term arising from the Jacobian is often omitted in practical computations. 
It is, however, important for the unitarity of the time evolution.
With a unitary time evolution the functional integral with weight distribution \eqref{eq:21}, \eqref{eq:33} constitutes a (non-relativistic) quantum field theory.
Our construction of a regularized functional integral as a probabilistic cellular automaton implements unitarity directly.

\subsection*{Expectation values}

\indent From $\tilde w[\sigma,\gamma]$ one can compute the expectation values of
arbitrary functionals $B[\sigma,\gamma]$ according to
\begin{equation}
\langle B[\sigma,\gamma]\rangle
=
\int \mathcal D \sigma\,\mathcal D \gamma\,
B[\sigma,\gamma]\,
\tilde w[\sigma,\gamma]\,.
\label{eq:31}
\end{equation}
They translate to time-ordered products of the corresponding operators evaluated with $\bar w[\sigma]$ or $\hat w[\gamma]$. 
For a proof we note that $\sigma(t)$ appears only in $\bar S(t-\varepsilon)$, and $\gamma(t)$ is present only in the product $\bar S(t)\bar S(t-\varepsilon)$. 
Consider the expression
\begin{align}
&\int \tilde{\mathcal D}\sigma(t)\,
\gamma(t)\,\bar S(t-\varepsilon)\nonumber\\
&=
\int
\tilde{\mathcal D}\sigma(t)
\tilde{\mathcal D}\sigma(t-\varepsilon)\!
\exp\Bigl\{
-i\Big(
\gamma(t)\sigma(t)
\nn \\
&\qquad-
\gamma(t-\varepsilon)\sigma(t-\varepsilon)
\Big)
\Bigr\}
\left(
-i\frac{\partial}{\partial\sigma(t)}
\right)
\hat S(t-\varepsilon)\,,
\label{eq:63C}
\end{align}
where we employ partial integration. 
For the computation of $\langle\gamma(t)\rangle$ one inserts into the product chain \eqref{eq:63A} for $\tilde w[\sigma,\gamma]$ a factor $\gamma(t)$.
This replaces effectively $\hat S(t-\varepsilon)$ by $-i\frac{\partial}{\partial\sigma(t)}\hat S(t-\varepsilon)$.
With,
\begin{align}
 &\int \tilde{\mathcal D}\gamma(t)\,
 \bar S(t)\bar S(t-\varepsilon) \nn \\
 =
 &\int \mathcal D\sigma(t-\varepsilon)\,
 \exp\{
    -i\left( \gamma(t-\varepsilon)\sigma(t-\varepsilon) \right)
 \} \nn \\
 &\times\hat S(t)\hat S(t-\varepsilon)\,
 \exp\{i\gamma(t-\varepsilon)\sigma(t-\varepsilon)\},
 \label{eq:63C1}
\end{align}
the functional integral over $\gamma$ for the product
\eqref{eq:63A} yields a product of factors $\hat S(t)$.
Multiplication with $\gamma(t)$ replaces in this product the factor
$\hat S(t-\varepsilon)$ by $-i\frac{\partial}{\partial\sigma(t)} \hat S(t-\varepsilon)$.
This corresponds precisely to the functional integral expression of the
expectation value of $\gamma(t)$ in terms of $\bar w[\sigma]$, as defined by eqs.~\eqref{eq:8}, \eqref{eq:9}.

\indent One can generalize this to functions $A(\gamma(t))$. 
For the expectation value $\langle A(\gamma(t))\rangle$ one replaces
$\hat S(t-\varepsilon)$ by $A\left(-i\frac{\partial}{\partial\sigma(t)}\right)\hat S(t-\varepsilon)$.
Consider next functions $B[\sigma,\gamma]$ which can be written in the form of a product
\begin{equation}
B[\sigma,\gamma]
=
\prod_t\tilde B(t)\,,
\label{eq:63D}
\end{equation}
where $\tilde B(t)$ depends either on $\gamma(t)$ or on $\sigma(t)$, but not on both.
For all factors $\tilde B(t)$ which are functions of $\gamma(t)$ one
replaces $\hat S(t-\varepsilon)$ by $\tilde B\left( -i\frac{\partial}{\partial\sigma(t)} \right)\hat S(t-\varepsilon)$.
For the factors which depend on $\sigma(t)$ one substitutes for $\hat S (t-\varepsilon)$ the expression $\tilde B\bigl(\sigma(t)\bigr)\hat S(t-\varepsilon)$.
We can use in this basis the standard operators $\hat\sigma(t)=\sigma(t)$, $\hat\gamma(t)=-i\frac{\partial}{\partial\sigma(t)}$.
As familiar in quantum mechanics, one can employ Heisenberg operators
acting on some common reference wave function or density matrix at a fixed $t=\bar t$. 
The functional integral expression \eqref{eq:31} for $B[\sigma,\gamma]$
corresponds then to time ordered products of Heisenberg operators, with larger $t$ on the left.

\indent Since the extended weight function $\tilde w[\sigma,\gamma]$ is complex, a real function $B[\sigma,\gamma]$ can result in a complex expectation value. 
Indeed, the time ordered product of hermitian operators is, in general, not a hermitian operator which would have a real expectation value. 
As an example, we may consider the function
\begin{equation}
B(t)
=
\frac12
\left(
C(t+\varepsilon)D(t)
+
C(t)D(t+\varepsilon)
\right)\,,
\label{eq:63E}
\end{equation}
where the operators $\hat C(t)$ and $\hat D(t)$ associated to $C(t)$ and $D(t)$ are hermitian. 
The expectation value reads
\begin{equation}
\langle B(t)\rangle
=
\frac12
\left\langle
\hat C(t+\varepsilon)\hat D(t)
+
\hat D(t+\varepsilon)\hat C(t)
\right\rangle
=
\langle\hat B(t)\rangle\,,
\label{eq:63F}
\end{equation}
where
\begin{align}
\hat B(t)
&=
\frac12
\left(
\hat C(t+\varepsilon)\hat D(t)
+
\hat D(t+\varepsilon)\hat C(t)
\right)\,, \nn \\
\hat B^\dagger(t)
&=
\hat B(t)
+
\frac12
\left(
\left[\hat D(t),\hat C(t+\varepsilon)\right]
+
\left[\hat C(t),\hat D(t+\varepsilon)\right]
\right)
\nonumber\\
&\approx
\hat B(t)
+
\frac{\varepsilon}{2}
\left(
\left[\hat D(t),\dot{\hat C}(t)\right]
-
\left[\dot{\hat D}(t),\hat C(t)\right]
\right)\,,
\label{eq:63G}
\end{align}
with hermitian operators
\begin{equation}
\dot{\hat C}(t)
=
i\left[H,\hat C(t)\right],
\quad
\dot{\hat D}(t)
=
i\left[H,\hat D(t)\right]\,.
\label{eq:63H}
\end{equation}
The term $\sim\varepsilon$ is antihermitian and therefore has a purely imaginary expectation value. 
In the continuum limit $\varepsilon\to0$ the operator $\hat B(t)$ becomes hermitian and $\langle B(t)\rangle$ real, given by the anticommutator of $\hat C$ and $\hat D$,
\begin{equation}
\langle B(t)\rangle
=
\left\langle
\frac12
\left\{
\hat C(t),\hat D(t)
\right\}
\right\rangle\,.
\label{eq:63I}
\end{equation}
These relations hold for a sufficiently smooth evolution such that expectation values  $\langle\hat{\dot{C}}(t)\hat D(t)\rangle$ etc. do not diverge as $\varepsilon^{-1}$ or stronger.
    
\indent A particular case are functions involving $\gamma$ and $\sigma$ at the same time $t$, as
\begin{equation}
B(t)
=
C\bigl(\sigma(t)\bigr)
A\bigl(\gamma(t)\bigr)\,.
\label{eq:63J}
\end{equation}
Similar to eq.~\eqref{eq:63C} one has
\begin{align}
&\int
\tilde{\mathcal D}\sigma(t)\,
C\bigl(\sigma(t)\bigr)
A\bigl(\gamma(t)\bigr)
\bar S(t-\varepsilon)
\nonumber\\
&=
\int
\tilde{\mathcal D}\sigma(t)
\tilde{\mathcal D}\sigma(t-\varepsilon)
\exp\{
-i[
\gamma(t)\sigma(t)
\nn \\
&-
\gamma(t-\varepsilon)\sigma(t-\varepsilon)
]
\}
A\left(
-i\frac{\partial}{\partial\sigma(t)}
\right)
\left[
C\bigl(\sigma(t)\bigr)
\hat S(t-\varepsilon)
\right]\,.
\label{eq:63K}
\end{align}
As a result, one replaces $\hat S(t-\varepsilon)$ by $A\bigl(\hat\gamma(t)\bigr)C\bigl(\hat\sigma(t)\bigr)\hat S(t-\varepsilon)$.
The operators are ordered with a factor of $\hat\gamma(t)$ on the left of the factors with $\hat\sigma(t)$.

\indent We can use these results for deriving the functional $B[\sigma,\gamma]$
which corresponds to some conserved quantity of the quantum system. 
As an example, we write $H^2$ in a form respecting this ordering
\begin{align}
H^2
&=
\hat\gamma_k\hat\gamma_\ell\,
F_k(\hat\sigma)F_\ell(\hat\sigma)
+
2i\,\hat\gamma_k
F_k(\hat\sigma)
\bigl(\partial_\sigma F\bigr)
\nonumber\\
&\quad
-\frac12
\frac{\partial}{\partial\sigma_k}
\left[
F_k\bigl(\partial_\sigma F\bigr)
\right]
+
\frac14
\bigl(\partial_\sigma F\bigr)^2\,.
\label{eq:63L}
\end{align}
where all operators are taken at the same $t$. 
This translates directly to a function $B[\sigma,\gamma]$ whose expectation value does not depend on time.

\indent For time-local observables depending on $\sigma(t)$ and $\gamma(t)$ at a given time $t$
the expectation values can also be computed from the extended wave function
$\varphi(t)$ by the quantum rule. The corresponding equal-time correlation
functions can be seen as classical correlation functions, with time-local
extended probability distribution $w(t,\sigma(t),\gamma(t)=|\varphi\bigl(t,\sigma(t),\gamma(t)\bigr)|^2$.
They have therefore to obey Bell's inequalities. These particular
correlations do not translate to the quantum correlations based on the
operators $\hat \sigma$ and $\hat \gamma$ for the subsystems corresponding
to $\bar w[\sigma]$ and $\hat w[\gamma]$. For these subsystems
simultaneously sharp values for $\sigma$ and $\gamma$ are not possible.
This is an example how classical correlations in an extended system are not accessible by the probabilistic information for a subsystem \cite{CWO1, CWPW}.

\section{Stochastic transport equations}

\indent We can include in the force $F$ a stochastic source $\eta$, $F(\sigma)\rightarrow F(\sigma)+\eta$.
Stochastic transport equations are used in
many areas of science, computations or economics.
Our quantum formalism includes this case. 
We find now a quantum system with a stochastic Hamiltonian involving $\eta$.
For a time-dependent stochastic source the Hamiltonian is time-dependent.

\indent The stochastic mean of an observable $B(\sigma,\gamma)$ is given by
\begin{align}
\bar B(\sigma,\gamma)
&=
\int\mathcal D\eta\,
P(\eta)\,
\left\langle
B(\sigma,\gamma;\,\eta)
\right\rangle
\nonumber\\
&=
\int
\mathcal D\eta\,
\mathcal D\sigma\,
\mathcal D\gamma\,
P(\eta)\,
\tilde w(\sigma,\gamma;\,\eta)\,
B(\sigma,\gamma)\,,
\label{eq:51}
\end{align}
with $P(\eta)$ the probability distribution for the stochastic source $\eta$. 
Probabilistic aspects arise here in two ways. 
There are first probabilistic initial conditions for $\sigma$, as encoded in the wave function $\varphi$ in eq.~\eqref{eq:21}. 
Second, the force is probabilistic due to the stochastic source $\eta$.
We focus on a Gaussian distribution $P(\eta)$,
\begin{equation}
P(\eta)
=
\mathcal N_\eta
\exp\left\{
-\int_t
\eta_k(t)\,
G_{k\ell}\,
\eta_\ell(t)
\right\}\,,
\label{eq:52}
\end{equation}
with $\mathcal N_\eta$ chosen such that $\int\mathcal D\eta\,P(\eta)=1$.
This results in
\begin{align}
&\bar B (\sigma, \gamma) = \tilde{\mathcal{N}} \int \mathcal{D}\sigma \mathcal{D}\gamma\, B(\sigma, \gamma) \varphi^\ast (t_f) \varphi(t_\text{in}) \exp(iS_S)\,,
\nn \\
&\exp\bigl(i S_S\bigr)
=
\int\mathcal D\eta\,
\exp\Bigg\{
\int_t
\Bigg[
i\gamma
\bigl(
\partial_t\sigma-F-\eta
\bigr)
\nn \\
&\hspace{3.5cm}-\eta_kG_{k\ell}\eta_\ell-\frac12\partial_\sigma F
\Bigg]
\Bigg\}\,.
\label{eq:53}
\end{align}

\indent The Gaussian integral over $\eta$ can be performed, resulting in
\begin{equation}
iS_S
=
iS_M
-
\frac12
\int_t
\gamma_k
G^{-1}_{k\ell}
\gamma_\ell
+
\text{const.}
\label{eq:54}
\end{equation}
The effect of the stochastic source is the addition to $S_M$ of a term quadratic in $\gamma$. 
In summary, the stochastic mean of the observable $B(\sigma,\gamma)$ reads
\begin{align}
\label{eq:55}
\bar B
&=
Z^{-1}
\int
\mathcal D\sigma
\mathcal D\gamma\,
B(\sigma,\gamma)\,
\varphi^*(t_f)\,
\varphi(t_{\mathrm{in}})
\\
&\quad\times
\exp\left\{
\int_t
\left[
i\gamma
\bigl(
\partial_t\sigma-F
\bigr)
-
\frac12
\gamma_kG^{-1}_{k\ell}\gamma_\ell
-
\frac12\partial_\sigma F
\right]
\right\}\,. \nn
\end{align}
This corresponds to the functional integral of refs.~[52--54], with probabilistic initial condition specified by $\varphi(t_{\mathrm{in}})$.
The boundary factor $\varphi^*(t_f)\,\varphi(t_{\mathrm{in}})$ is often set to a constant for practical computations.
The normalization factor $Z^{-1}$ obtains by setting $B(\sigma,\gamma)=1$, $\bar B=1$.
For observables depending only on $\sigma$ one can further perform the Gaussian integration over $\gamma$.

\indent Quantum systems with a stochastic Hamiltonian can be viewed as an ensemble of quantum systems for each value of $\eta$. 
Since the Hamiltonian depends on $\eta$, an ensemble average typically does not admit a periodic behavior.
This contrasts with the quantum system for a given fixed time-independent $\eta$. 
On the other hand, if an observable is time-independent for every value of $\eta$, this also holds for the ensemble average.
Examples are statistical observables related to symmetry transformations,
as quantum momentum or angular momentum in case of translation or rotation symmetry in space, 
or conserved charges corresponding to ``internal'' symmetry transformations acting on components of $\sigma(\vec x)$.

\section{Conclusions}

\indent This note focusses on the conceptual interpretation of probabilistic transport equations.
We argue that these systems are genuine quantum systems, with a Schr\"odinger equation for a complex wave function and non-commuting operators for observables.
This has far reaching implications: Quantum systems can emerge as a particular case of classical probabilistic systems.
The quantum systems for probabilistic transport equations provide counter-examples for no go theorems claiming that an embedding of quantum mechanics in classical probabilistic physics is not possible. 
Some of the assumptions of such theorems are simply not met in our setting.
For example, Bell's inequalities for the correlation of two observables $A$ and $B$ assume the existence of simultaneous probabilities for pairs $A_i$ and $B_j$ of possible measurement values for these observables.
Such a ``classical correlation function'' simply does not exist for statistical observables since they do not have fixed values for the ``microstates'' or ``configurations''.

\indent Once the quantum structure of probabilistic transport equations is established, one may ask about its practical use.
One avenue is the use of the statistical observables. 
Conserved quantities as the quantum angular momentum for rotation invariant systems, quantum momentum for translation invariant systems, or conserved charges constrain the dynamics.
Another interesting issue is the focus on quantum subsystems with low-dimensional wave functions. 
The characteristic quantum effects as  interference and the uncertainty relation are expected to be particularly prominent in this case.
The quantum structure reveals characteristic properties, as the existence of periodic probability distributions in case of time-independent $F(\sigma)$.

\indent For stochastic transport equations one finds stochastic quantum mechanics with a Hamiltonian depending on a stochastic force. 
This washes out some of the characteristic quantum properties as the existence of periodic probability distributions.
Also the number of conserved statistical observables is reduced. 
One may ask what is new beyond the well established MSRJD-functional integral \cite{MSR,JAN,DOM} for stochastic transport equations.
On the conceptual side it is the recognition of the quantum structure and the association of the ``response field'' $\gamma$ with a statistical observable related to the field-derivative of the wave function.
On the practical side, conserved statistical observables constrain the dynamics. 
This may be used for an improved understanding.

\nocite{*}
\bibliography{refs}

\end{document}